\documentclass[10pt]{iopart}
\usepackage[T1]{fontenc}
\usepackage{ae}
\usepackage{aecompl}
\usepackage{graphicx}
\usepackage{fixltx2e}
\usepackage{calc}
\usepackage{nag}
\usepackage[UKenglish]{babel}

\usepackage[dvips]{hyperref}
\hypersetup{pdftitle={Algebraically contractible topological tensor network states},
pdfauthor={S. J. Denny et al.},
bookmarksnumbered,
colorlinks=true,
pdfborder={0 0 0},
linkcolor=blue,
citecolor=blue}

\newcommand{\question}[1]{\relax\marginpar{\relax\textcolor{red}{[#1]}}}

\newcommand{\remove}[1]{#1}
\renewcommand{\remove}[1]{}
\renewcommand{\question}[1]{}

\newcommand{\half}{\mbox{$\textstyle\frac{1}{2}$}}
\newcommand{\ket}[1]{\mbox{$|#1\rangle$}}

\newcommand{\bra}[1]{\mbox{$\langle #1|$}}
\newcommand{\braket}[2]{\mbox{$\left\langle #1|#2\right\rangle$}}
\newcommand{\brakets}[2]{\mbox{$\langle #1|#2\rangle$}}

\newcommand{\proj}[1]{\ket{#1}\bra{#1}}
\newcommand{\av}[1]{\langle #1\rangle}

\newcommand{\eqrs}[2]{equations~(\ref{#1}--\ref{#2})}
\newcommand{\fir}[1]{figure~\ref{#1}}
\newcommand{\Fir}[1]{Figure~\ref{#1}}
\providecommand{\Sref}[1]{Section~\ref{#1}}
\providecommand{\eref}[1]{(\ref{#1})}

\providecommand{\text}[1]{{\rm #1}}

\newcommand{\mini}[2]{\begin{minipage}[c]{#1}\centering #2\end{minipage}}
\newcommand{\minigrpx}[2]{\mini{#1}{\includegraphics{#2}}}

\newcommand{\sixj}{F^{ijm}_{kln}}
\providecommand{\mod}{{\rm mod\ }}
\newcommand{\COPY}{$\bullet$}
\newcommand{\XOR}{$\oplus$}

\begin{document}

\title{Algebraically contractible topological tensor network states}

\author{S~J~Denny$^1$, J~D~Biamonte$^2$, D~Jaksch$^{1,2}$ and S~R~Clark$^{2,1}$}
\address{$^1$Clarendon Laboratory, University of Oxford, Parks
Road, Oxford OX1 3PU, United Kingdom}
\address{$^2$Centre for Quantum Technologies, National University of
Singapore, 3 Science Drive 2, Singapore 117543}
\ead{s.denny1@physics.ox.ac.uk}

\date{\today}

\begin{abstract}
We adapt the bialgebra and Hopf relations to expose internal structure in the ground state of a Hamiltonian with $Z_2$ topological order.  
Its tensor network description allows for exact contraction through simple diagrammatic rewrite rules.  
The contraction property does not depend on specifics such as geometry, but rather originates from the non-trivial algebraic properties of the constituent tensors.  
We then generalise the resulting tensor network from a spin-\half{} lattice to a class of exactly contractible states on spin-$S$ degrees of freedom,  yielding the most efficient tensor network description of finite Abelian lattice gauge theories.  
We gain a new perspective on these states as examples of two-dimensional quantum states with algebraically contractible tensor network representations. 
The introduction of local perturbations to the network is shown to reduce the von Neumann entropy of string-like regions, creating an unentangled sub-system within the bulk in a certain limit.  
We also show how perturbations induce finite-range correlations in this system.  
This class of tensor networks is readily translated onto any lattice, and we differentiate between the physical consequences of bipartite and non-bipartite lattices on the properties of the corresponding quantum states.  
We explicitly show this on the hexagonal, square, kagome and triangular lattices.
\end{abstract}


\maketitle

% \section*{To do.}
% \begin{enumerate}
%   \item \st{Incorporate citation to Tagliacozza} \cite{PhysRevB.83.115127}
%   \item Fix up buerschaper complaint
%   \item \st{Fix typo}
% \end{enumerate}

\section{Introduction}
The inherent difficulty in describing quantum many-body systems is a consequence of the ``curse of dimensionality'' --- the cost of describing a state grows exponentially with the number of degrees of freedom.  Progress has been made on this front, exploiting results from quantum information theory which constrain physically relevant states to a small corner of Hilbert space.  In one dimension the problem has been effectively solved for short-range interactions with matrix product states (MPS)~\cite{PhysRevLett.75.3537} and the density matrix renormalisation group (DMRG)~\cite{PhysRevLett.69.2863,RevModPhys.77.259,nishino1999density,PhysRevLett.102.057202}, but  in higher dimensions DMRG becomes computationally unfeasible.  To tackle this, projected entangled pair states (PEPS)/tensor network states (TNS)~\cite{nishino2001twodimensional,PhysRevA.81.052338,verstraete2008matrix,corboz2010simulation} directly generalise MPS, while tree tensor networks (TTN)~\cite{PhysRevA.74.022320} and the multi-scale entanglement renormalisation ansatz (MERA)~\cite{PhysRevLett.101.110501,2009arXiv0912.1651V,PhysRevB.80.165129,PhysRevA.81.010303} instead make direct use of a hierarchical structured tensor network.

Tensor networks are capable of describing quantum states ranging from those with short-range entanglement (such as symmetry-breaking states) up to topological phases of matter.  However, while the tensor network approach is highly versatile in terms of the quantum states it can describe, its utility is threatened by the difficulty of contracting the tensor network to obtain reduced density matrices and expectation values.  Approximate methods exist to perform this contraction.  PEPS were first introduced along with an algorithm based on interpreting the network contraction in terms of MPS and matrix product operators (MPOs)~\cite{2004cond.mat..7066V}. The tensor renormalisation group (TRG)~\cite{PhysRevLett.99.120601,PhysRevB.78.205116} method has also been presented as a means for performing this contraction approximately in polynomial time, and has been demonstrated to provide an accuracy beginning to approach that of quantum Monte Carlo (QMC) methods where applicable~\cite{PhysRevB.81.174411,PhysRevLett.106.127202}.  
Other methods to efficiently contract tensor networks include corner transfer matrix methods~\cite{PhysRevB.80.094403} and a coarse-graining method explicitly including and exploiting a local symmetry for computational gain~\cite{PhysRevB.83.115127}.

By contrast, our interest here is in developing a class of quantum states having TNS descriptions which are exactly contractible through simple diagrammatic rewrite rules~\cite{lafont1992penrose,lafont1995equational}.
Already it is known that MERA forms an efficiently-contractible class of PEPS~\cite{PhysRevLett.105.010502}, and that certain states (such as the MERA representation of the toric code) are fixed points under an entanglement renormalisation procedure~\cite{PhysRevLett.100.070404,PhysRevB.79.195123}.
However, the simplicity of our TNS contraction does not rely on the underlying network topology, on tensors constrained to be unitaries or isometries, or on being a fixed point of an abstract renormalisation procedure. Rather, these network rewrite properties are graphical rules which encode algebraic properties of the tensors.  Furthermore, these tensor networks can readily be defined on any lattice, allowing us to consider  how the states represented by the tensor network vary with lattice geometry (in arbitrary spatial dimensionality) and the local on-site dimension.

The states we consider display topological order.  In \Sref{sec:topological} we show that they are the deconfined phases of Abelian lattice gauge theories.  The fractional quantum Hall effect~\cite{PhysRevLett.58.1252} provided the first examples of topologically ordered phases.  Topological phases of matter are also found in lattice gauge theories~\cite{RevModPhys.78.17}, spin liquids~\cite{anderson1987resonating} and topological insulators~\cite{kane2008condensed}.  Interest in these exotic phases of matter stems from their novelty of lying outside the conventional Landau symmetry-breaking paradigm.  They are characterised by their long-range entanglement and topology-dependent ground state degeneracy.  Their study is further motivated by the potential application of topological phases of lattice models to fault-tolerant quantum computers and quantum memories.  Kitaev's toric code~\cite{kitaev2003fault} provided the first example of a topological quantum error-correcting code based on the $Z_2$ group.  Our class of states is a subclass of those accessible within Levin and Wen's string-net formalism~\cite{PhysRevB.71.045110}, and as such is already known to have tensor network representations which are fixed points of the TRG procedure~\cite{PhysRevB.79.085118,PhysRevB.79.085119}.  We provide a nascent and exemplary class of states, algebraically contractible via a considerably simpler route.

\section{Summary of results}
In the next section, % \secr{sec:boolean}
we introduce tensor network states in a general setting, along with the special class of boolean tensors introduced in~\cite{2010arXiv1012.0531B} and the pertinent rules regarding their manipulation.  These are illustrated by the well known GHZ state.  In \Sref{sec:topological} we briefly review models with topological order, then introduce $Z_2$ lattice gauge theory with a discussion of its physics as a case study.  We refer to the ``$Z_2$ state'' as the ground state of a Hamiltonian with $Z_2$ topological order.  In \Sref{sec:contract_Z2} we reproduce the tensor network state description of the $Z_2$ state from~\cite{PhysRevB.79.085118} (adapted for the hexagonal lattice) and present a new property of this TNS, namely that it is exactly contractible through the bialgebra, Hopf and tensor fusion relations introduced in \Sref{sec:boolean}.  We use our construction to calculate the two-point correlation functions of this state, and confirm that it has a zero correlation length.  We also show a direct calculation of the topological entanglement entropy from the TNS.  In \Sref{sec:excitations} we discuss the excitations in $Z_2$ lattice gauge theory, and demonstrate how these can be included using ``impurity'' tensors in the network.  The exact contractibility property enables the network to be contracted around the impurities provided their number is finite.  We demonstrate this and consider string-like regions of the lattice containing a variable number of impurity tensors, calculating the reduced density matrix and von Neumann entropy for this region.  As a second application, we calculate the topological entanglement entropy in the region around an impurity tensor.

\Sref{sec:generalisation} takes the $Z_2$ TNS and generalises it to a class of tensor networks representing the deconfined phases of all finite Abelian lattice gauge theories.  This is done in a way in which the tensors used to construct this class obey the same rules that enabled the exact contraction in the $Z_2$ case.  We also show that these are, in a certain sense, the most efficient tensor network representations possible of these quantum states that can be expected.  In \Sref{sec:dimensionality} we translate these tensor networks onto square, kagome and triangular lattices and compare the properties of the corresponding quantum states.  We find that on bipartite lattices the network will contract by precisely the same mechanism, with identical mathematical structures pointing to identical physics in every case.  By contrast, on non-bipartite lattices the network remains contractible but with some important differences.  In particular, these tensor networks are seen to represent non-trivial quantum states, but can only possess $Z_2^n$ topological order.  \Sref{sec:conclusion} concludes the paper.

\section{Tensor network states and boolean algebra}\label{sec:boolean}
A tensor network provides an efficient means to describe a quantum many-body system of spins living on a lattice (or any system which may be mapped as such). 
Product states through to symmetry-breaking and topological phases are all accessible within this description.  We use the physical lattice geometry to create a network in which tensors are associated to the physical spins.  For each spin with coordination number $n$, the corresponding tensor network description incorporates an order ($n$+1) tensor $A^{m_i}_{\alpha\beta\gamma\ldots}$ at each vertex, with $n$ internal indices $\alpha,\beta,\gamma\ldots$ providing the network edges and one physical index $m_i$ representing the state of the spin in a physical basis.  All internal indices are summed over in any network.  Typically, internal indices are $d$-dimensional and the physical indices are $D$-dimensional, however in this paper we will always consider the specific case $d=D$.  On a hexagonal lattice in which the spins are three-fold coordinated, a general tensor network state is expressed as 
\begin{equation}\label{eq:general_TNS}
\ket{\Psi} =
\sum_{\{m_i\}=0}^{D-1}\sum_{\{\alpha,\beta,\gamma\ldots\}=0}^{d-1} A^{m_1}_{\alpha\beta\gamma}A^{m_2}_{\alpha\delta\epsilon}A^{m_3}_{\beta\zeta\eta}A^{m_4}_{\gamma\theta\iota}\ldots \ket{\{m^{}_i\}}.
\end{equation}

The notation \ket{\{m^{}_i\}} indicates a quantum state of all spins on the lattice in which the $i^\text{th}$ physical spin takes the value $m_i$ in the chosen basis.  
\Fir{fig:TNS} gives a graphical representation of the quantum state in \eref{eq:general_TNS}.  In the figure, the square blocks represent tensors, open legs represent the physical indices, and lines connecting tensors denote the sums in \eref{eq:general_TNS} --- \emph{contraction} over pairs of indices.  Periodic boundary conditions are typically employed on a sufficiently large network, so that the internal legs on the boundaries wrap around the lattice.  In addition, requirements of translational and rotational invariance can often restrict the tensors to all be identical and rotationally symmetric, respectively, allowing for a highly compact state description.  However, evaluating any probability amplitude or other property of such a state, such as reduced density matrices, involves contraction of a network.  In general this is an unfeasible operation.  To carry out the contraction naively, e.g.~by contracting the tensors along lines in one direction first has exponentially large memory requirements.  This is seen in extracting a single amplitude from a TNS on a square lattice: each tensor has $d^4D$ components and a line of $n$ tensors contracted together forms a tensor with $d^{2n+2}D^{n}$ components.  This is prohibitive for all but the smallest of lattices.  Numerical schemes to approximately perform this contraction exist (such as the MPS-MPO method~\cite{2004cond.mat..7066V} and TRG~\cite{PhysRevLett.99.120601,PhysRevB.79.085118}) but here our interest focusses on specific cases in which we can carry out the full contraction analytically for lattices of arbitrary size.

\begin{figure}[htbp]
\centering
\includegraphics{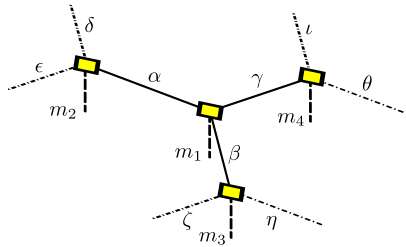}
\caption{Graphical representation of a section of a general tensor network state with amplitudes $\Psi\left(\left\{ m_i \right\} \right) = \sum_{\{\alpha,\beta,\gamma\ldots\}=0}^{d-1} A^{m_1}_{\alpha\beta\gamma}A^{m_2}_{\alpha\delta\epsilon}A^{m_3}_{\beta\zeta\eta}A^{m_4}_{\gamma\theta\iota}\ldots$, in which the physical degrees of freedom are arranged on the vertices of a hexagonal lattice.  The square blocks represent tensors, the lines between them indicate summations over pairs of internal indices, and open (long-dashed) legs correspond to physical indices.}
\label{fig:TNS}
\end{figure}
We now briefly review the ``constituent network components'' introduced in~\cite{2010arXiv1012.0531B} which will be the building blocks of our class of algebraically contractible states.  We restrict the dimensionality of all tensor indices to $d=2$ as relevant for spin-\half{} systems, and furthermore all tensor components to be 0 or 1.  This is the realm of the so-called ``boolean tensors''~\cite{2010arXiv1012.0531B,lafont2003towards}.  We introduce two tensors, COPY (\COPY) and XOR (\XOR),
\begin{equation}\label{eq:COPY_XOR_defn}
\begin{tabular}{cc}
COPY & XOR \\
\minigrpx{3cm}{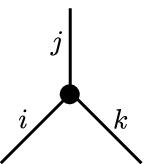} & \minigrpx{6cm}{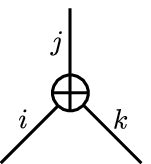}
\end{tabular}
\end{equation}
\begin{equation*}
\delta_{ijk}=\left\{
\begin{array}{cll}
1, & i = j = k,\\
0, & \text{otherwise,}
\end{array}
\right.
\quad
X_{ijk}=\left\{
\begin{array}{cll}
1, & i + j + k = 0\ {\left( \mod 2 \right),}\\
0, & \text{otherwise.}
\end{array}
\right.
\end{equation*}
The legs of these graphical representations signify their indices.  
The COPY tensor defined above is a three-index generalisation of Kronecker's delta, and its extension to an arbitrary number of legs is straightforward.  
The second major component, the XOR tensor is related to the COPY tensor~\cite{2010arXiv1010.4840B}.  
We can write it in terms of COPY through the contraction of each leg with a $d=2$ Fourier matrix $H_{jk} = (-1)^{jk}$, denoted by  \minigrpx{0.3cm}{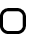},
\begin{equation}\label{eq:PLUS_COPY_link}
\minigrpx{2.4cm}{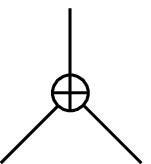} = \minigrpx{2.4cm}{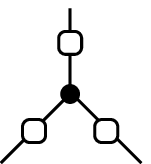}.
\end{equation}

We can likewise define $n$-legged XOR tensors in terms of $n$-legged COPY tensors.  The XOR tensors have a number of useful diagrammatic properties.  They are clearly fully symmetric under the exchange of legs.  They have units, i.e.~basis vectors which when contracted with any leg of the tensor give an identity matrix
\begin{equation}
\minigrpx{2cm}{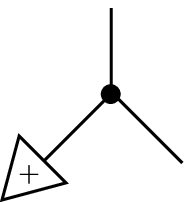} = \minigrpx{1.5cm}{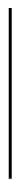}, \quad \minigrpx{3cm}{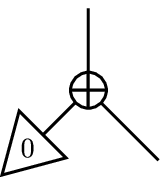} = \minigrpx{1.5cm}{equation4b},
\end{equation}
where $\ket{+} = (1\ 1)^T$ and $\ket{0} = (1\ 0)^T$.  These same vectors form copy-points with the two tensors
\begin{equation}
\minigrpx{2cm}{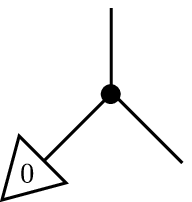} = \minigrpx{2cm}{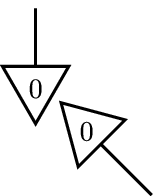}, \quad \minigrpx{2cm}{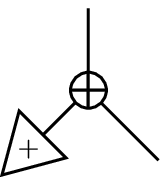} = \minigrpx{2cm}{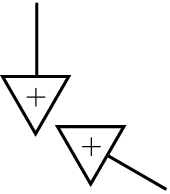}.
\end{equation}
The COPY tensor obeys a so-called ``tensor fusion rule''\footnote{This is \emph{not} related to the fusion rules in topological quantum field theory, which we will refer to exclusively as ``branching rules.''} allowing a network of COPY tensors to be amalgamated at will, e.g.
\begin{equation}
\minigrpx{2.5cm}{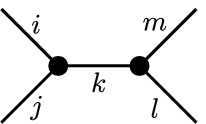} = \minigrpx{1.5cm}{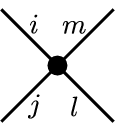},
\end{equation}
providing a purely diagrammatic interpretation for the equation $\sum_k \delta_{ijk} \delta_{klm} = \delta_{ijlm}$.  This is also true of the XOR tensor,\footnote{Later, we will consider a generalisation of this tensor to higher dimensions.  Note that, unlike COPY, this generalised XOR only obeys a fusion rule if it is constructed from a direct sum of $Z_2$ groups.}
\begin{equation}
\minigrpx{2.5cm}{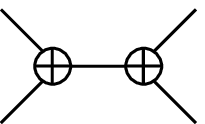} = \minigrpx{2cm}{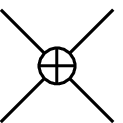}.
\end{equation}
The COPY and XOR tensors together obey a bialgebra law~\cite{kassel1995quantumch3,kock2004frobenius},
\begin{equation}\label{eq:bialgebra}
\minigrpx{3cm}{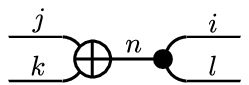} = \minigrpx{4cm}{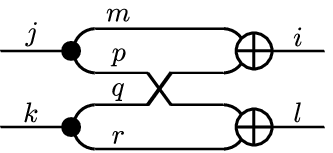}.
\end{equation}
Later on, we will consider a generalised tensor $B_{ijk}$ while maintaining the requirement that it obeys a bialgebra law with a higher dimensional COPY tensor.  
Expressing the bialgebra law~\eref{eq:bialgebra} as a tensor equation,
\begin{equation}
\sum_n  B_{jkn}\delta_{lin} = \sum_{mpqr} \delta_{jpm}\delta_{krq}B_{mqi}B_{prl},\label{eq:bialgebra_condition}
\end{equation}
we derive a condition that these generalised tensors $B_{ijk}$ must obey to satisfy a bialgebra law with COPY.  All components $B_{jkl} =$ 0 or 1, and further $B_{jki}B_{jkl} = 0$ if $i \neq l$.  These place a heavy restriction on the number and location of the non-zero elements of $B_{ijk}$.
Finally, the COPY and XOR tensors also obey a Hopf law~\cite{lafont2003towards,kock2004frobenius}
\begin{equation}\label{eq:hopf}
\minigrpx{2.7cm}{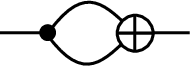} = \minigrpx{2.7cm}{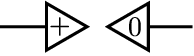}.
\end{equation}

We emphasise that these rewrite rules are defined up to an overall complex constant of proportionality, and that detailed proofs can be found in \cite{2010arXiv1012.0531B}.  As a simple application of this graphical formalism, we can express the $n$-party GHZ state \ket{\Psi_\text{GHZ}} as
\begin{eqnarray}
\ket{\Psi_\text{GHZ}} &= 2^{-\frac{1}{2}}\left( \ket{00\ldots 0} + \ket{11\ldots 1} \right)\label{eq:GHZ}\nonumber\\
 &= \minigrpx{7cm}{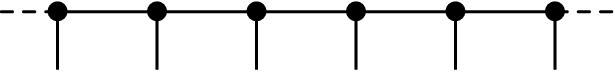}.\label{eq:GHZ_pic}
\end{eqnarray}
It is straightforward to see that this represents \ket{\Psi_\text{GHZ}}:  contracting each physical leg with \ket{0} or \ket{1}, we see that the coefficient is non-zero only when \ket{00\ldots 0} and \ket{11\ldots 1} are inputted.  
We can calculate any two-site reduced density matrix by taking the complex conjugate of the state (\bra{\Psi_\text{GHZ}}) and contracting over all but two physical legs.  In the chosen basis all amplitudes are real, and so the complex conjugate may be overlooked.  Our convention is that upwards- and downwards-directed physical legs correspond to bras and kets respectively.  The contraction of the state follows directly from the COPY tensor fusion rule.
\begin{eqnarray}
\rho_{AB} &= \minigrpx{7cm}{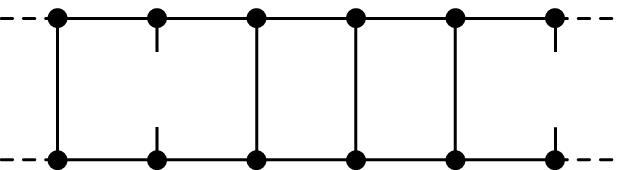}\nonumber\\
&= \minigrpx{5.5cm}{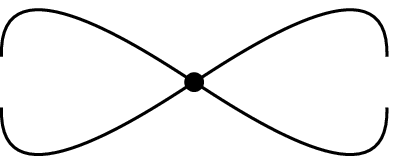}\nonumber\\
&= \qquad\frac{1}{2}\left( \proj{00} + \proj{11} \right).\label{eq:rho_GHZ}
\end{eqnarray}
As noted above, the rewrites are only defined up to proportionality, so in \eref{eq:rho_GHZ} we have normalised at the final stage.  Diagrammatically it is clear that $\rho$ takes this form regardless of which sites we choose, showing that the correlation length for this state is infinite.  

\section{Topological models}\label{sec:topological}
Before introducing the tensor network state for the $Z_2$ model and generalisations, we provide a brief introduction to the main aspects of Levin and Wen's class of string-nets~\cite{PhysRevB.71.045110}, as our resulting quantum states will form a subclass of these models.  A state in this class consists of a set of labels on the edges of a hexagonal lattice.  The labels correspond to directed string types, and the Hilbert space is composed of all possible combinations of these labels on each edge.  A particular string-net condensate is completely specified by a set of ``data''~\cite{PhysRevB.71.045110}. 
\begin{enumerate}
\item A list of directed string types, $i=\left(0,\ldots, N-1\right)$.  Each string type has a ``conjugate'' type --- the same string with the reverse orientation, which is written as $i^*=j$.  Undirected strings have $i^*=i$.
\item A branching rule tensor $T_{ijk} = 1$ if string types $i$, $j$ and $k$ may meet at a vertex, $0$ otherwise.  We will use $T$ to denote the branching rules tensor throughout, but in the $Z_2$ case $T_{ijk}=X_{ijk}$.
\item A quantum dimension $d_i$ for each string type.  The total quantum dimension $\mathcal{D}$ is defined as $\mathcal{D} = \left(\sum_i d_i^2\right)^{1/2}$.
\item A six-index tensor $\sixj$ describing the recoupling relations between different string types.  These are identities allowing for the amplitudes of locally different string-net configurations to be related to one another~\cite{PhysRevB.71.045110}, viz.
\begin{equation}
\Psi\left(\minigrpx{1.9cm}{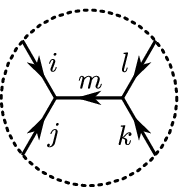}\right) = \sum_n\ F^{ijm}_{kln}\ \Psi\left(\minigrpx{1.9cm}{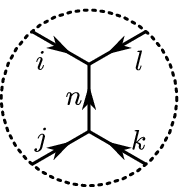}\right).
\end{equation}
In this notation, $\Psi\left(\minigrpx{.35cm}{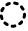}\right)$ is the string-net amplitude for a configuration $\minigrpx{.35cm}{circ}$  on the lattice.  The piece displayed above is a local section of a string-net, the remainder is arbitrary and equal on both sides of the equation above.  $\sixj$ satisfies the so-called ``pentagon equation'' to guarantee a single-valued wavefunction~\cite{turaev1994quantum}.
\begin{equation}\label{eq:pentagon}
\sum_n F^{mlq}_{kp^*n} F^{jip}_{mns^*} F^{js^*n}_{lkr^*} = F^{jip}_{q^*kr^*}F^{riq^*}_{mls^*}
\end{equation}
\end{enumerate}
Many of the components of $\sixj$ are related by symmetry.
The $F$-tensors contain the branching rules implicitly, for example
\begin{equation}\label{eq:6j_branching}
T^{}_{ijk}=\frac{v_iv_j}{v_k}F^{ijk}_{j^*i^*0}, \quad v_i^2 = d^{}_i.
\end{equation}
Finding solutions to \eref{eq:pentagon} is difficult in general, however it is the case that all groups and quantum groups generate a solution in which the string type index runs over the irreducible representations of the group and $\sixj$ is its corresponding $6j$-symbol~\cite{PhysRevB.71.045110}.

We now focus on $Z_2$ lattice gauge theory as the simplest example of a topological quantum field theory.  In terms of spin-\half{} degrees of freedom living on the sites of a kagome lattice, its Hamiltonian is
\begin{equation}\label{eq:full_Z2_Hamiltonian}
H=U\hspace{-10pt} \sum_{\text{vertices\ } v}\left(1-\prod_{i \in v} \sigma_i^z\right)-g \hspace{-13pt}\sum_{\text{plaquettes\ } p}\ \prod_{i \in p} \sigma_i^x - J\hspace{-2pt}\sum_{\text{edges\ }i}\sigma^z_i.
\end{equation}
The terms ``vertices'', ``plaquettes'' and ``edges'' refer to those on the dual (hexagonal) lattice.
We can also discuss this Hamiltonian using string language.  The model is equivalent to one type of undirected string living on the edges of a hexagonal lattice.  In terms of the eigenstates of $\sigma_z$, \ket{0} and \ket{1}, we interpret the state \ket{1} on an edge as having a string present, and \ket{0} being empty.  The strings correspond to the ``electric flux'' lines of the lattice gauge theory.  
Generally, we also assume $g\ge 0$, $J\ge 0$.  This model is dual to an Ising model in (2+1) dimensions (in which the strings are domain boundaries), and contains a quantum phase transition~\cite{PhysRevB.79.085118}.  When $g=0$, the ground state is ferromagnetic with all sites spin-up.  This is the confined phase of the lattice gauge theory, and in string language the ground state is absent of strings.  As $g/J$ increases past a critical value $(g/J)_c$, the ground state becomes a weighted superposition of closed string configurations.
In the limit $J=0$, the above Hamiltonian \eref{eq:full_Z2_Hamiltonian} becomes exactly solvable,
\begin{equation}
H=-\hspace{-10pt}\sum_{\text{vertices\ } v}\ \prod_{i \in v} \sigma_i^z
-\hspace{-13pt}\sum_{\text{plaquettes\ } p}\ \prod_{i \in p} \sigma_i^x. 
\end{equation}
Provided that the ratio $g/U$ is finite, it does not affect the ground state, and so we set both $U$ and $g$ to 1 for simplicity.  We focus on the ground state in this limit,  the ``$Z_2$ state'' which is an \emph{equal} superposition of all configurations with closed strings,
\begin{equation}
\ket{\Psi} = \sum_{S \text{\ closed}}\ket{S},
\end{equation}
where $S$ is a basis configuration of the lattice, and `$S$ closed' signifies the subset of such configurations in which there are only closed loops --- no open or branching strings.  This is the deconfined phase of the lattice gauge theory.  Despite its simplicity, this state is topologically ordered.  As a final remark, the $Z_2$ state is a stabiliser state, since it is an eigenstate of the set of stabilisers $\prod_{i \in v} \sigma_i^z$ with eigenvalue $+1$.

\section{Algebraic contraction of the \texorpdfstring{$Z_2$}{Z2} state}\label{sec:contract_Z2}
\begin{figure}[tbp]
\centering
\includegraphics{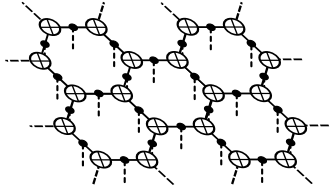}
\caption{Graphical depiction of the tensor network for the $Z_2$ state, from \cite{PhysRevB.79.085118}.  The network components are the COPY tensor $\delta_{\alpha\beta\gamma}$ (\COPY) and the XOR tensor $X_{\alpha\beta\gamma}$ (\XOR).  The XOR tensors enforce the ``closed-loop'' constraint of the $Z_2$ state.}
\label{fig:Z2_state}
\end{figure}
With the basic physics of $Z_2$ lattice gauge theory now presented, we turn to a tensor network representation of an ideal $Z_2$ state~\cite{PhysRevB.79.085118}.
This tensor network is depicted in \Fir{fig:Z2_state}, and can now be seen to be composed of precisely the COPY and XOR tensors we have introduced in \Sref{sec:boolean}.
Intuitively, the COPY tensors transmit the chosen basis states (\ket{0}, \ket{1}) on the physical indices to the internal indices of the XOR tensors, which then enforce the branching rules (closed-loop constraint).
We now demonstrate how the graphical rewrite rules introduced in \Sref{sec:boolean} allow us to contract this state exactly.
First consider an area in which all physical indices are contracted.
The double-layered network structure results from contracting physical indices on two copies of the state (from \bra{\Psi} and \ket{\Psi}),
\begin{equation}\label{eq:double_lattice}
\braket{\Psi}{\Psi} = \minigrpx{4.5cm}{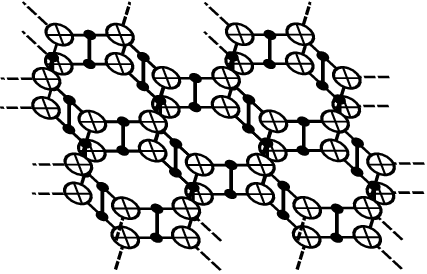}.
\end{equation}
Focussing on a small region of the lattice, the diagrammatic local rules previously introduced are put to use.
We manipulate the network, using the fusion rule to rearrange the COPY tensors followed by the bialgebra law.
\begin{eqnarray}
\minigrpx{2.7cm}{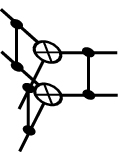} &= \minigrpx{2.7cm}{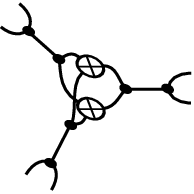}\label{eq:flatten_hex_start}\nonumber\\
&= \minigrpx{2.7cm}{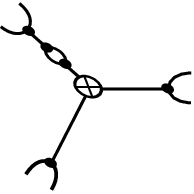}\nonumber\\
&= \minigrpx{2.7cm}{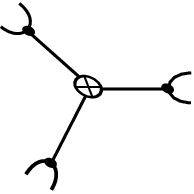}\label{eq:flatten_hex_end}
\end{eqnarray}
Consequently, the double layer network flattens to a single layer and can subsequently be contracted via the XOR fusion rule,
\begin{equation}\label{eq:contracted_lattice}
\minigrpx{4.5cm}{equation19} = \minigrpx{4.5cm}{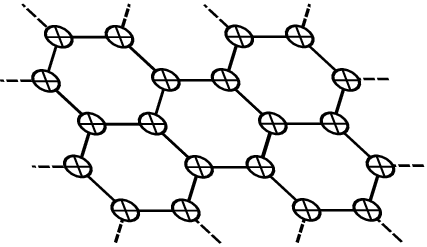}.
\end{equation}

We can use these rewrite rules to calculate quantities of interest, such as two-site reduced density matrices and the topological entanglement entropy.  In a way analogous to that for the GHZ state, the method readily reveals the nature of the correlations in this state. Specifically the two-site reduced density matrix is given by a double-layer network similar to that above, but with the two sites of interest left uncontracted.
\begin{eqnarray}
&\minigrpx{8cm}{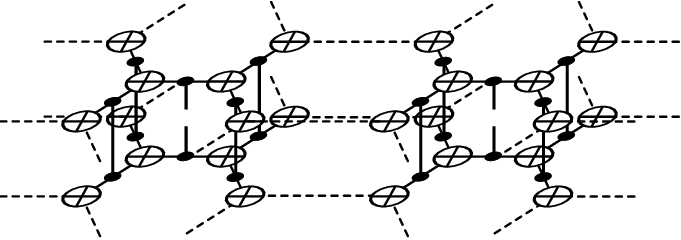}\label{eq:two_site_rho_lattice}\nonumber\\
&=\minigrpx{7.5cm}{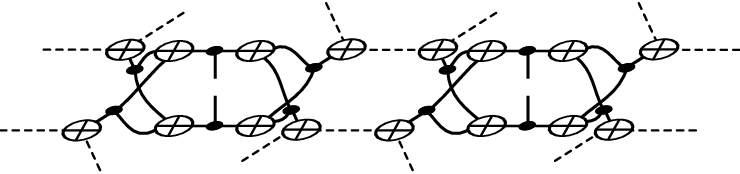}\nonumber\\
&=\minigrpx{5cm}{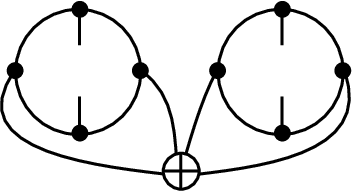}\nonumber\\
&=\minigrpx{5cm}{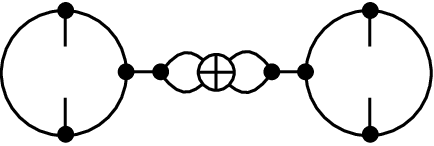}\nonumber\\
&=\minigrpx{5cm}{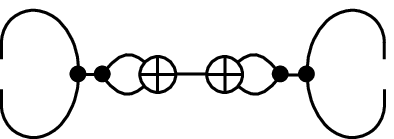}\nonumber\\
&=\minigrpx{5cm}{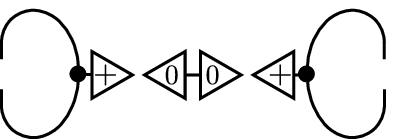}\quad{\text{(via\ the\ Hopf\ law)}}\nonumber\\
&=\minigrpx{5cm}{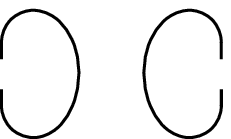}\nonumber\\
&=\mini{5cm}{$\frac{1}{4}\left(\proj{0} + \proj{1}\right)^{\otimes 2}$}\label{eq:two_site_rho_analytic}
\end{eqnarray}

The local rewrites clearly expose the product state nature of the two-site reduced density operator.  Consider the connected correlation functions $\av{O(\bi{r}_1)O(\bi{r}_2)}_c = \av{O(\bi{r}_1)O(\bi{r}_2)} - \av{O(\bi{r}_1)}\av{O(\bi{r}_2)}$, in which $\bi{r}_i$ are positions on the lattice, and $O$ is any local operator.  Then  $\av{O(\bi{r})} = \frac{1}{2}\Tr O$ and $\av{O(\bi{r}_1)O(\bi{r}_2)} = \frac{1}{4}\left(\Tr O \right)^2$, hence $\av{O(\bi{r}_1)O(\bi{r}_2)}_c = 0$.  As such, the correlation length of this state is zero.  This is also true of product states, yet this state is characterised by its pattern of long-range entanglement as revealed by the topological entanglement entropy~\cite{kitaev2006topological}.
We can employ a similar strategy to calculate this quantity directly.  The entropy of entanglement $S$ is defined as the von Neumann entropy of a reduced state $\rho_A$ of a bipartite system $AB$,
\begin{equation}
S = -\Tr \left( \rho_A \log \rho_A \right), \qquad \rho_A = \Tr_B \left( \proj{\Psi}\right).
\end{equation}
Generally, the entropy of entanglement of a region in a topological phase scales with the region's boundary $L$ as $S = \alpha L - \gamma + \ldots$, with terms which vanish as $L \rightarrow \infty$ omitted~\cite{kitaev2006topological,2006PhRvL..96k0405L}.  As the state has a correlation length of zero, we have precisely that  $S = \alpha L - \gamma$.  Here $\gamma$ is a subleading constant term which arises in the case of topological phases ($\gamma>0$) \cite{kitaev2006topological}. 
On a lattice, a region enclosing $N$ sites has an ill-defined boundary $L$, so a scheme based on differences is employed to ensure the cancellation of the leading term. The Kitaev-Preskill topological entanglement entropy~\cite{kitaev2006topological} is defined as  $S_\text{top} = S_A + S_B + S_C - S_{AB} - S_{BC} - S_{CA} + S_{ABC} = -\gamma$, with $A$, $B$ and $C$ denoting mutually adjacent subregions of the lattice, such that the composite region $A \cup B \cup C$ is simply connected as shown in \fir{fig:S_top_connectivity}.
\begin{figure}[htbp]
\centering
\includegraphics{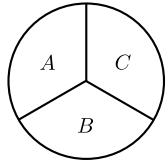}
\caption{Connectivity of regions as taken when calculating the topological entanglement entropy $S_\text{top} = S_A + S_B + S_C - S_{AB} - S_{BC} - S_{CA} + S_{ABC}$.  The regions and definition of $S_\text{top}$ are chosen such that the term in $S$ scaling with the boundary $L$ cancels in $S_\text{top}$.}
\label{fig:S_top_connectivity}
\end{figure}

The zero correlation length of these states enables a simple calculation of the topological entanglement entropy, since the limit $L\rightarrow \infty$ need not be taken.  The leading term $\alpha L$ is cancelled in the construction, and all other potential $L$-dependent terms are zero.  Consequently, we may proceed by considering just three adjacent sites.
From the reduced density matrices above, $S_A = \log 2$, $S_{AB} = 2\log 2$.  The three-site reduced density matrix $\rho_{ABC}$ evaluates to
\begin{equation}\label{eq:rho_three_site}
\minigrpx{3cm}{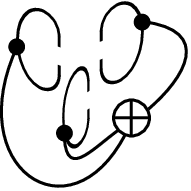}:\quad \rho_{ABC} = \sum_{ijk} X_{ijk}\proj{ijk}
\end{equation}
This reduced density matrix does not factorise and is diagonal with $2^2$ equal non-zero entries. It gives $S_{ABC} = 2\log 2$ yielding the topological entanglement entropy $S_\text{top} = -\log 2$ in complete agreement with~\cite{PhysRevA.71.022315}, but computed here by very different means.

\section{Excitations in \texorpdfstring{$Z_2$}{Z2} lattice gauge theory}\label{sec:excitations}
The $Z_2$ model has four kinds of elementary quasiparticle excitation operators: trivial, ``magnetic flux'', ``electric charge'' and ``electric/magnetic bound states''.  The first three have bosonic statistics, while the last is fermionic.  These operators act along strings on the lattice, creating pairs of quasiparticles, one at each end of the string.  The excitations are gapped (with energy of order $U$, as in \eref{eq:full_Z2_Hamiltonian}).

Within our formalism, an excitation may be included by inserting an ``impurity tensor'' (\minigrpx{0.3cm}{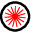}) on a site which enforces an open string there.
The rest of the lattice may be contracted around any impurities, leaving a network sufficiently small to contract directly. 
As an example, we calculate the density matrix of a string of $n=4$ spins as shown in \fir{fig:lattice_string}.  The ``open'' tensors (\minigrpx{0.3cm}{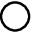})  are unspecified, and will be set shortly to be either standard $Z_2$ or impurity tensors. 

\begin{figure}[htbp]
\centering
\includegraphics{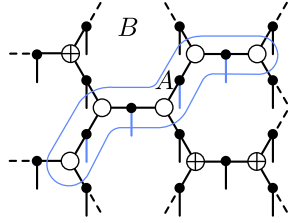}
\caption{Local perturbation to the ground state via the inclusion of impurity tensors.  The hexagonal lattice is divided into a string-like region $A$, and the remainder region $B$.}
\label{fig:lattice_string}
\end{figure}
A contraction over the degrees of freedom in the bulk (region $B$) yields the density matrix for region $A$:
\begin{equation}\label{eq:rho_string}
\rho_A\quad = \minigrpx{5cm}{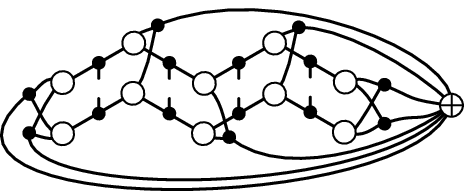}
\end{equation}
We can choose the impurity tensors to be intermediate between the extremes of open and closed loops, where $\theta$ parametrises this set,  
\begin{equation}\label{eq:impurity_T_defn}
X'_{ijk}=\left\{
\begin{array}{cll}
\cos^{\frac{1}{2}} \theta,\ & i+j+k = 0\ (\mod 2), \\
\sin^{\frac{1}{2}} \theta,\ & i+j+k = 1\ (\mod 2),
\end{array}
\right. \quad \left(i,j,k\right)=0,1.
\end{equation}
Setting $\theta = 0$ corresponds to the standard $Z_2$ state, while $\theta = \pi/2$ corresponds to enforcing open strings at the location of each impurity tensor.  We can now set each of the $(n+1)$ tensors in \eref{eq:rho_string} to be either the standard $Z_2$ tensor or impurity $X'$ tensor, and examine the variation of the von Neumann entropy with $\theta$ and the number of impurities $m$.  This is plotted in \fir{fig:string_S}. 
\begin{figure}[tbp]
\center
\includegraphics{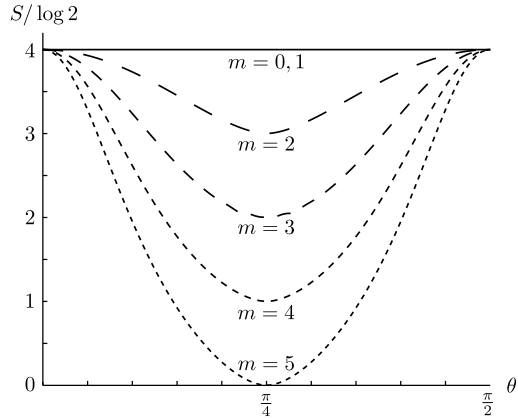}
\caption{Von Neumann entropy $S$ for the four spins in a string-like region of the lattice, in the configuration shown in  \fir{fig:lattice_string}.  Each curve corresponds to varying the impurity tensor parameter $\theta$ for a string with $m$ of the five vertex tensors along the string set to the impurity $X'_{ijk}$, and the rest as the standard $Z_2$ $X_{ijk}$.  The impurity tensor $X'$ is defined in \eref{eq:impurity_T_defn} such that $\theta=0$ corresponds to the standard $Z_2$ $X$-tensor, $\theta=\pi/2$ corresponds to enforcing open strings and $\theta=\pi/4$ as the mid-point between these two extremes is an equal superposition of the two, and has a factorisation as in \eref{eq:T_product}.  Note, $S$ depends only on the number of $X'$ tensors and the value of $\theta$.  It is independent of the relative positioning of these tensors.}
\label{fig:string_S}
\end{figure}

Physically, if all of the vertices in the string are standard $Z_2$ tensors, the reduced density matrix is proportional to the identity and so $S$ takes its maximum value, $S = n \log 2$.  A second intuitive limit is to choose $\theta = \pi/4$ and set all $(n+1)$ tensors along the string to be the impurity $X'_{ijk}$.  For this value of $\theta$, the impurity tensor factorises,
\begin{equation}\label{eq:T_product}
\minigrpx{3cm}{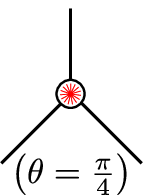} = \minigrpx{3cm}{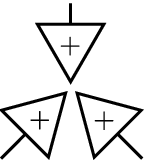}.
\end{equation}
Substituting this into the density matrix \eref{eq:rho_string} immediately yields
\begin{eqnarray}
\rho_{\ (\theta=\pi/4)} \quad &= \minigrpx{5.5cm}{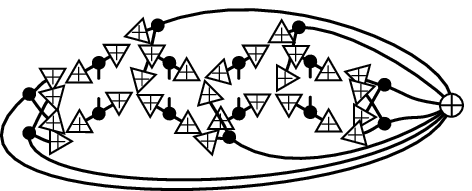} \nonumber\\
 &= \minigrpx{5cm}{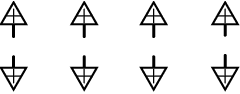} \nonumber\\
 &= \mini{5cm}{\proj{++++}}.
\end{eqnarray}
This four-spin subsystem is in a pure state --- it is an island, unentangled with the remainder of the topological phase, a fact not easily revealed by other means.

For general $m$, $(n+1-m)\log 2 \le S \le n \log 2$ for $1 \le m \le n+1$.  For $m=0$ there are no impurity tensors and so no $\theta$-dependence.  The form of the curves can be understood as follows:  if all tensors along the string are set to $X'$, $\theta$ varies the subsystem between the standard $Z_2$ and a pure state.  If instead $n < m -1 $ tensors are set to $X'$, provided that they are all adjacent it is possible to trace out the ends of the string leaving a shorter string entirely consisting of $X'$ tensors.  The degrees of freedom traced out were responsible for the entanglement and the residual non-zero $S$ even at $\theta=\pi/4$.  If instead the $X'$ tensors are not adjacent, this procedure cannot be carried out.  
Nevertheless \eref{eq:rho_string} indicates that the moving the impurity tensor only rearranges the tensor components of $\rho$, and does so in such a way that $S$ remains unchanged.

As a second application, we can make use of the algebraic contraction property to study the topological entanglement entropy for the composite region $ABC$ shown in \fir{fig:lattice_excitation_pair} as a function of $\theta$.  This configuration consists of two impurity tensors placed arbitrarily on the lattice, and the regions $A$, $B$ and $C$ each containing one of the spins adjacent to the first impurity tensor.  The overlap of this state with the ``ideal'' $Z_2$ state is $\cos\theta$.  The tensor network for $\rho_{ABC}$ contracts to
\begin{eqnarray}\label{eq:rho_ABC_excitation}
\rho_{ABC}\quad&=\minigrpx{3.5cm}{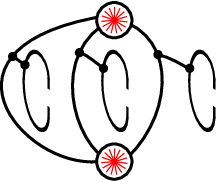}\\
&= \frac{1}{4} \text{\ diag} \left( \cos^2\theta, \sin^2 \theta, \cos^2\theta, \ldots ,\sin^2 \theta\right). \nonumber
\end{eqnarray}
The one- and two-site reduced density matrices are proportional to the identity as before.  Consequently, if we calculate the topological entanglement entropy as before, considering just these three sites, we find
\begin{equation}\label{eq:S_top}
S_\text{top}=-\log 2 -\cos^2\theta\log\left(\cos^2\theta\right)-\sin^2\theta\log\left(\sin^2\theta\right).
\end{equation}
When we expand the region $ABC$ to include more spins, we find \eref{eq:S_top} remains unchanged \emph{until} the second excitation tensor is contained within the region $ABC$.  
At this point, we retrieve the result $S_\text{top}=-\log 2$ as required for this phase.  
Thus, the excitation tensors introduce correlations between the local spins in the two regions, and this must be taken into account in the regions when calculating the topological entanglement entropy in this way.
At both $\theta = 0$ and $\theta = \pi/2$ the three-site calculation gives the full value ($-\log 2$) for the state; in the case of a `pure' excitation, the no correlations are introduced because we have effectively just locally relabelled the spins.  At the intermediate value $\theta = \pi/4$, the correlations are maximal.  At this value of $\theta$, $X'_{ijk}$ factorises to a product as in \eref{eq:T_product}.

If we place the second excitation tensor at the vertex immediately adjacent to the spin in region $A$ (say), we have the situation considered in the first example and see that only $S_A$, $S_{AB}$ and $S_{AC}$ vary with $\theta$.  These do not cancel in $S_\text{top}$, leading to the $\theta$-dependence in \eref{eq:S_top}.  If instead the second excitation tensor is not immediately adjacent to any of the spins in the region $ABC$, only $S_{ABC}$ varies, but precisely as required to yield the same result for $S_\text{top}$.

\begin{figure}[htbp]
\centering
\includegraphics{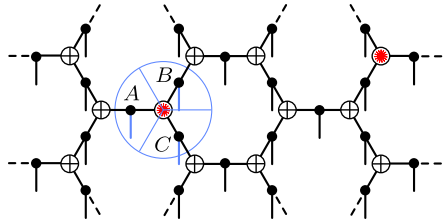}
\caption{Accessing excited states via the inclusion of impurity tensors.  The two red starred vertices signify the impurity tensors, while the rest of the network is the standard $Z_2$ state  (\fir{fig:Z2_state}).  The partition of regions around one of the impurities is marked:  the regions $A$, $B$ and $C$ contain one spin each, all immediately adjacent to an excitation tensor.}
\label{fig:lattice_excitation_pair}
\end{figure}
Our formalism allows for straightforward computations concerning localised perturbations and excitations, and their effect on the von Neumann entropy of string-like regions as in \fir{fig:lattice_string} and the topological entropy as in \fir{fig:lattice_excitation_pair}.  These are two examples of the physical quantities which are directly calculable by this method; the physics concerning such configurations of isolated excitations is not otherwise easily accessible.

\section{Generalisation to a class of contractible states}\label{sec:generalisation}
We have shown that the $Z_2$ state for a spin-\half{} system has an efficient tensor network description which can be contracted efficiently by local rewrite rules to evaluate quantities of interest.  We now generalise this network to provide an efficient description of all finite Abelian lattice gauge theories for spin-$S$ systems.  We retain precisely the tensor network structure introduced in \fir{fig:Z2_state}, but instead of using the labels $\{0,1\}$ for the indices, we now replace them with the elements of the $Z_2$ group, $\{ I,e \}$ with $e^2 = I$.  We can equivalently define the COPY tensor as
\begin{equation}\label{eq:COPY_defn_group}
\delta_{\alpha\beta\gamma}=\left\{
\begin{array}{cll}
1 & \alpha = \beta = \gamma,\\
0 & \text{otherwise.}
\end{array}
\right. \quad \left(\alpha,\beta,\gamma\right)\in \{I,e\}
\end{equation}
In the case of the XOR tensor defined in \eref{eq:COPY_XOR_defn}, we make use of the group product.  The branching rules $T_{\alpha\beta\gamma}$ may be equivalently redefined as 
\begin{equation}\label{eq:XOR_defn_group}
T_{\alpha\beta\gamma}=\left\{
\begin{array}{cll}
1 & \alpha\beta\gamma = I,\\
0 & \text{otherwise.}
\end{array}
\right. \quad \left(\alpha,\beta,\gamma\right)\in \{I,e\}
\end{equation}

We continue to represent $T_{\alpha\beta\gamma}$ graphically with \XOR.  The generalisation to all finite Abelian groups is straightforward, and proceeds by replacing $Z_2$ with a group $G$, for which $(\alpha,\beta,\gamma) \in G$ in the definitions above.  We have restricted focus to Abelian groups, since in order to connect these to their string-net condensate descriptions, we will make use of the one-one correspondence between an Abelian group's elements and its irreducible representations~\cite{serre1977linear}.  In the case that $G \cong Z_d$, $T_{\alpha\beta\gamma}$ is a very intuitive generalisation of XOR, namely a `PLUS' tensor as introduced in~\cite{2010arXiv1010.4840B}.  A PLUS tensor with $d$-dimensional indices has coefficient 1 when its indices sum to 0 $(\mod d)$.  In this case, it is connected to a $d$-dimensional COPY tensor through Fourier matrices as in \eref{eq:PLUS_COPY_link}, with components $H_{jk} = e^{i2\pi jk/d}$, where the integers $j$ and $k$ label the elements of the cyclic group.
\begin{equation}\label{eq:PLUS_COPY_link_general}
T_{\alpha\beta\gamma} \quad= \minigrpx{2.4cm}{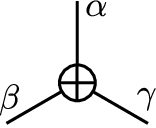} = \minigrpx{2.4cm}{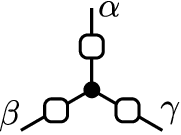} = \minigrpx{2.4cm}{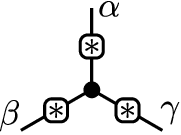}
\end{equation}
Diagrammatically, we represent $H$ and its complex conjugate $H^*$ with \minigrpx{0.3cm}{H} and \minigrpx{0.3cm}{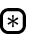} respectively.  In the finite Abelian case, $G$ can be decomposed as the direct sum $G\cong Z_{n_1} \oplus Z_{n_2} \oplus \ldots \oplus Z_{n_k}$, in which $n_i$ are powers of primes. \Eref{eq:PLUS_COPY_link_general} applies with $H$ taking a tensor product structure $H = H_{n_1} \otimes H_{n_2} \otimes \ldots \otimes H_{n_k}$ \footnote{If instead we had based the tensor network around a non-Abelian group, this property would not hold and we expect the contraction of the network, if still possible, to be more involved.}.  At this point, we have constructed a tensor network with an identical structure to before but with the tensors themselves generalised.
When attached to a physical index, the COPY tensor can be thought of as converting the chosen basis states \ket{i} to group elements on the internal indices of the $T$ tensors. 
The COPY tensor sets the dimension of the physical degrees of freedom to be $ d=\left|G\right|$, the order of the group, i.e.~this is a spin $S=\frac{1}{2}\left(|G|-1\right)$ state.

We want these newly-defined tensors to be contractible by similar means to the $Z_2$ network.  Our key requirements are $(i)$ a connection of the $T$ tensors to the COPY tensor through $H$, so that we can use the COPY tensor fusion rule to contract the bulk, and $(ii)$ a bialgebra law to flatten the lattice.  The first of these has been demonstrated in \eref{eq:PLUS_COPY_link_general}, and the second holds as long as $T_{\alpha\beta\gamma}$ defined in \eref{eq:XOR_defn_group} satisfy the condition \eref{eq:bialgebra_condition} required for a tensor to obey the bialgebra relation \eref{eq:bialgebra} with COPY.  That this is true can be shown as follows.  First, all elements $T_{\alpha\beta\gamma}=0$ or 1.  Second, the product of two elements $T_{\alpha\beta\gamma}T_{\alpha\beta\delta}=0$, since $T_{\alpha\beta\gamma}=1$ implies $\alpha\beta\gamma=I$.  The group structure guarantees the existence of unique inverses, so it follows that $T_{\alpha\beta\gamma}T_{\alpha\beta\delta}=0$: if $\alpha\beta\gamma=I$, $\alpha\beta\delta\neq I$ unless $\gamma=\delta$. Then the condition is satisfied.

Consequently, the contraction performed in \eqrs{eq:double_lattice}{eq:two_site_rho_lattice} proceeds just as in the case $d=2$, with only a slight alteration after the flattening step.  In general, it is no longer true that $H$ is self-inverse.  However, $H$ in \eref{eq:PLUS_COPY_link_general} can be freely replaced with $H^*$ and it is a property of Fourier matrices that $HH^*=I$.  Consequently, provided the lattice is bipartite, $T_{\alpha\beta\gamma}$ may be decomposed into COPY~$\times\ H$ and COPY $\times\ H^*$ on the two sublattices, respectively,
\begin{eqnarray}
\minigrpx{4cm}{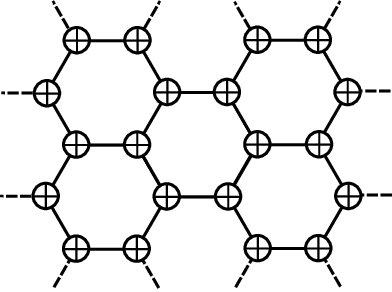} &= \minigrpx{4cm}{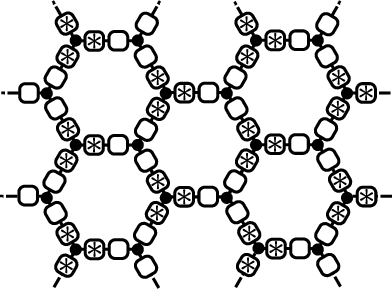} \label{eq:H_Hs_decomposition_hex}\nonumber\\
&= \minigrpx{4cm}{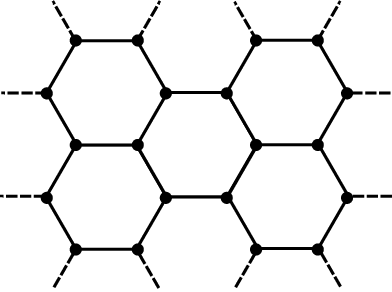}.
\end{eqnarray}
This enables the tensor fusion rule to contract the bulk state.  
We have therefore readily generalised the $Z_2$ state to a class of exactly contractible states based on finite Abelian groups.  We can make some immediate observations about the properties of these quantum states.  They are non-trivial as the $T$ tensors exclude any configuration not satisfying the branching rules and further, they consist of an equal superposition of allowed configurations owing to the equality of all non-zero coefficients in the network.

It just remains to show that these exactly contractible networks represent the deconfined phases of Abelian lattice gauge theories.  To do this we construct the string-net representation of these phases.  The $6j$-symbol $\sixj$\ is defined only if the four triads of labels $\{i,j,m\}$, $\{k,l,m^*\}$, $\{i,n,l\}$ and $\{j,k,n^*\}$ are allowed by the branching rules.  The $6j$ symbols of Abelian groups have all equal coefficients, therefore $\sixj=1$ if the triplets above are allowed, and for convenience we set $\sixj=0$ otherwise.  For completeness we note that these are the ground states of a very general class of exactly solvable Hamiltonians~$H$~\cite{PhysRevB.71.045110}, acting on $d$-dimensional spins on the edges of a lattice
\begin{equation}\label{eq:string_net_hamiltonian}
H=-\hspace{-5pt}\sum_{\text{vertices\ }v} Q_v - \hspace{-8pt}\sum_{\text{plaquettes\ } p} B_p,\quad B_p=\frac{1}{d^2}\sum_{s=0}^N B_p^s.
\end{equation}
Here $Q_v$ is a projector on a lattice site, projecting onto the subspace of allowed string configurations at that vertex,
\begin{equation}
Q_v\ \Psi\left(\minigrpx{1.4cm}{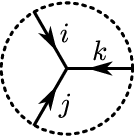}\right) = T_{ijk}\ \Psi\left(\minigrpx{1.4cm}{equation36}\right),
\end{equation}
and $B^s_p$ acts on plaquettes $p$ in a rather involved manner,
\begin{eqnarray}
B^s_p\ \Psi\left(\minigrpx{1.4cm}{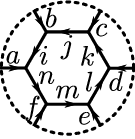}\right) = \nonumber\\
\hspace{0pt}\sum_{\bar \imath \bar \jmath \bar k \bar l \bar m \bar n}\hspace{0pt} F^{an^*i}_{s^*\bar \imath \bar n^*} F^{bi^*j}_{s^*\bar \jmath \bar \imath^*} F^{cj^*k}_{s^*\bar k \bar \jmath^*} F^{dk^*l}_{s^*\bar l \bar k^*}F^{el^*m}_{s^*\bar m \bar l^*} F^{fm^*n}_{s^*\bar n \bar m^*} \Psi\left(\minigrpx{1.4cm}{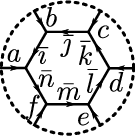}\right).
\end{eqnarray}
Note that $B^s_p$ has a much simpler intuitive interpretation~\cite{PhysRevB.71.045110}, namely it acts on $\Psi$ to introduce a loop of type $s$ around a plaquette $p$.  This can then be incorporated into the lattice via the recoupling relations to give the form above.  

The branching rules can be extracted from the $6j$-symbols via \eref{eq:6j_branching}, and these are identical to the branching rules of our state because of the correspondence between the group's elements and its irreducible representations.  The equality of the $6j$-symbol components implies we can transform one string-net configuration into another with no alteration of the coefficient, thus the ground state consists of an equal superposition of all configurations consistent with the branching rules.  This is precisely our tensor network state defined above.

Finally for these states, we can calculate directly the topological entanglement entropy.  The calculation proceeds as before, and again we find the three-site reduced density matrix $\rho_{ABC}= \sum_{\alpha\beta\gamma} T_{\alpha\beta\gamma}\proj{\alpha\beta\gamma}$, $\{\alpha,\beta,\gamma\}=(1\ldots d)$.  From this and \eref{eq:XOR_defn_group}, the one- and two-site density matrices are again seen to be of product form.  Combining these results, $S_\text{top} = -\log d$ for an Abelian lattice gauge theory with $d=\left|G\right|$.  This is well-known: Abelian anyons have quantum dimension $d_i=1$, so for total quantum dimension $\mathcal{D}$,  $\mathcal{D}^2$ is just the number of superselection sectors $=\left| G \right|^2$, e.g.~4 for $Z_2$ gauge theory.  It is also known that the total quantum dimension relates to the topological entanglement entropy via $S_\text{top} = -\log \mathcal{D} =-\log \left| G \right|$~\cite{kitaev2006topological}.  However, it is demonstrated here using a very different method.

We have taken the $Z_2$ tensor network and shown that it naturally generalises to a TNS class describing the deconfined phases of Abelian lattice gauge theories.  This TNS class also represents these states as efficiently as possible.  A reduced density matrix $\rho_D$ of a single spin having Hilbert space dimension $D$, has a von Neumann entropy which is bounded by $0 \le S \le \log D$.  Since an Abelian group of order $|G|$ generates a lattice gauge theory with a one-site reduced density matrix having $S = \log |G|$, it is necessary that to reproduce the local physical properties of this state the tensors must have internal legs of dimension $D = d \ge |G|$.  As our TNS class has internal indices with dimension $d=|G|$, it is therefore the most efficient representation of these quantum states in the sense that the internal bond dimension in the network is minimised.   

\section{Dimensionality and coordination}\label{sec:dimensionality}
We have presented an efficient tensor network description of the deconfined phases of finite Abelian gauge theories on a hexagonal lattice.  Through our construction, we can readily vary the lattice geometry on which the tensor network is based and determine the consequences for the underlying quantum state.  No changes to the method are required on bipartite lattices, but for clarity we demonstrate contraction on the square lattice.  The overlap \brakets{\Psi}{\Psi} exemplifies the contraction of the double-layer network,
\begin{equation}
\braket{\Psi}{\Psi} = \minigrpx{5.4cm}{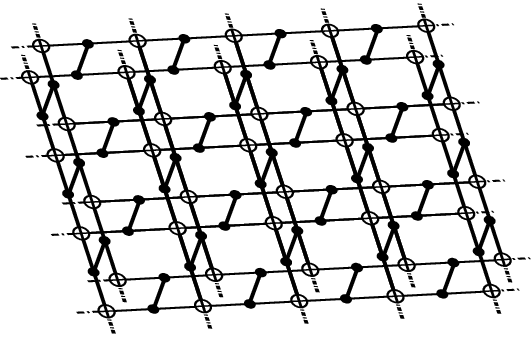}.
\end{equation}
Focussing on a small region of the lattice, the diagrammatic rewrite rules are put to use again.  For a tensor with components all in the set $\{0,1\}$ it is clear that (c.f. \eref{eq:flatten_hex_start})
\begin{equation}
\minigrpx{4cm}{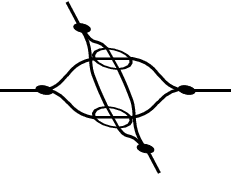}=\minigrpx{4cm}{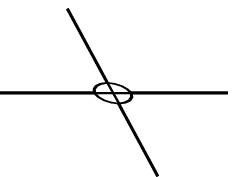}.
\end{equation}
Formally, this follows from rearrangements through the bialgebra relation.  The network flattens, and we can use the alternating decomposition into $H$ and $H^*$ 
just as in \eref{eq:H_Hs_decomposition_hex} followed by the COPY fusion rule.
\begin{equation}
\Rightarrow \braket{\Psi}{\Psi} \rightarrow \minigrpx{5.4cm}{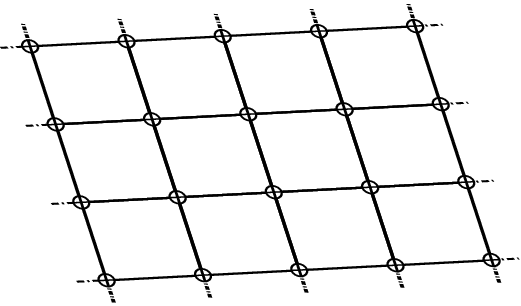}.
\end{equation}
That we can handle the tensor network here in exactly the same manner as on the hexagonal lattice demonstrates that this class of states possesses identical physics on these two lattices, and indeed on any other bipartite lattice such as the simple cubic lattice in three dimensions.  In summary, this is well known, but demonstrated here in a particularly clean and simple way.

We can also construct these tensor networks on non-bipartite lattices (e.g.~kagome and triangular).  We expect this to take us outside the string-net class, as the orientation convention for the branching rules~\cite{PhysRevB.71.045110} cannot be satisfied on all vertices of a non-bipartite lattice.  The states remain exactly contractible, but demonstrating this requires more algebraic rewrites.  On the kagome lattice, we first ``flatten'' the double-layer network just as before.  
Taking inspiration from an application of the TRG routine to treat the kagome lattice~\cite{PhysRevB.78.205116}, we can rewrite the network as follows,
\begin{eqnarray}
\minigrpx{4cm}{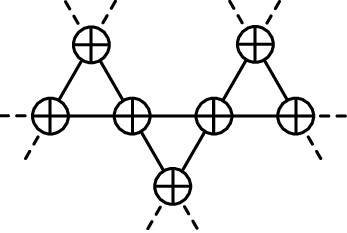} &= \minigrpx{4cm}{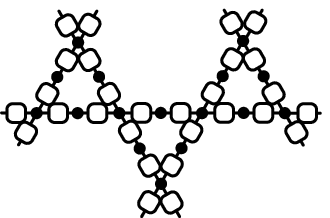}\nonumber\\
&= \minigrpx{4cm}{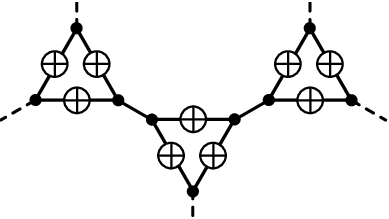}\ .
\end{eqnarray}
Each PLUS tensor has been factored into a COPY tensor and Fourier matrices.  We are free to insert two-legged COPY tensors, and combine these with the Fourier matrices to give two-legged PLUS tensors.  We can then analyse the resulting triangular structures using group theoretic notions.  
Graphically we denote this structure as a white dot,
\begin{equation}\label{eq:white_COPY}
S_{\alpha\beta\gamma}\quad = \minigrpx{3cm}{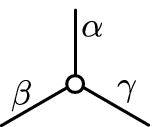} = \minigrpx{3cm}{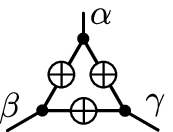}.
\end{equation}
To determine its properties, we input three group elements $\alpha,\beta,\gamma \in G$ into the right-hand side of \eref{eq:white_COPY}.  This tensor element $S_{\alpha\beta\gamma}=1$ if $\alpha=\beta^{-1}$, $\beta=\gamma^{-1}$ and $\gamma=\alpha^{-1}$ simultaneously. This implies that all indices must be equal, $\alpha=\beta=\gamma$, and also that they must be self-inverse, $\alpha=\alpha^{-1}$.  Otherwise $S_{\alpha\beta\gamma}=0$.  Hence this structure behaves as a COPY tensor but only for self-inverse group elements.  We will refer to it as a ``sub-COPY tensor.''  This tensor obeys a fusion rule both with itself and with the COPY tensor,
\begin{equation}
\minigrpx{2.5cm}{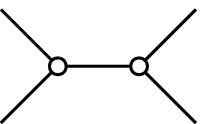} = \minigrpx{2.5cm}{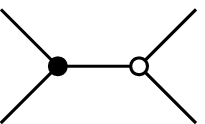} = \minigrpx{2 cm}{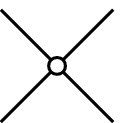}.
\end{equation}
This allows for similar contractions to those previously considered.  A particular difference arises in the topological entanglement entropy.  For the kagome lattice, the four-site reduced density matrix is diagrammatically expressed as
\begin{equation}\label{eq:S_top_kagome}
\rho_{ABCD}\quad = \minigrpx{3.5cm}{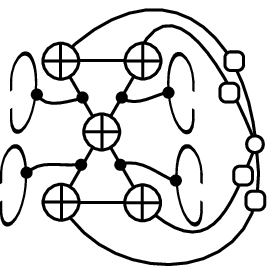}.
\end{equation}
This can be simplified further in special cases, in particular if $G \cong Z_2^n$ (a direct sum of $n$ $Z_2$ groups), then $\circ = \bullet$ and the $T$ tensors can be combined via the fusion rule to give a four-site density matrix with the structure of \eref{eq:rho_three_site}.  Otherwise, the effect of the sub-COPY tensor is that only self-inverse elements (equivalently, $Z_2$ subgroups) make a contribution to $S_\text{top}$.  By this, we mean that if $G\cong Z_2 \oplus Z_3 \oplus Z_4$, we would calculate $S_\text{top}=-\log(2\times 2)$, as the $Z_2$ group and $Z_2$ subgroup in $Z_4$ contribute, but the $Z_3$ group does not.  
This is entirely consistent with the expected complications arising from a simple translation of tensor network states for string-nets to non-bipartite lattices.  The orientation convention for the branching rules cannot be satisfied, and this will have consequences for states containing directed strings.
The exception to this is when $G \cong Z_2^n$, in which case all strings are undirected and the full topological order is present.  
This does not preclude the possibility of constructing such theories on non-bipartite lattices, merely that the single-bond tensor network presented here is insufficient.
Nevertheless, a state in our class on the kagome lattice constructed from the $Z_3$ group (for definiteness) is in this case not topologically ordered ($S_\text{top}=0$), yet is an example of a non-trivial spin-1 quantum state with exact contraction properties as previously discussed.  This can be compared with a state constructed from the $Z_3$ group on a square lattice, which \emph{does} yield a topologically ordered spin-1 quantum state.  In both cases, each PLUS tensor enforces the same correlations in four nearby spins and it is then the network connectivity which sets whether or not this induces topological order in the corresponding quantum state.

Turning finally to the triangular lattice, a similar result occurs.  The network can be rearranged as follows, yielding a COPY and sub-COPY tensor network which contracts to the same result as on the kagome lattice.
\begin{eqnarray}
\minigrpx{4cm}{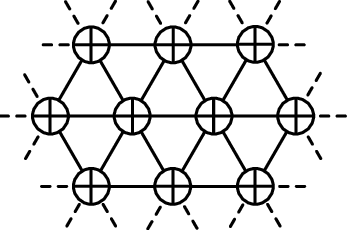} &= \minigrpx{4cm}{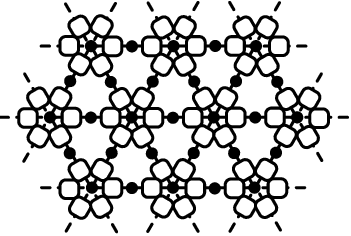}\nonumber\\
&= \minigrpx{4cm}{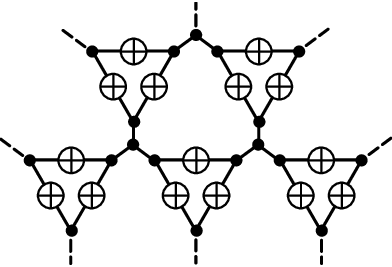}\nonumber\\
&= \minigrpx{4cm}{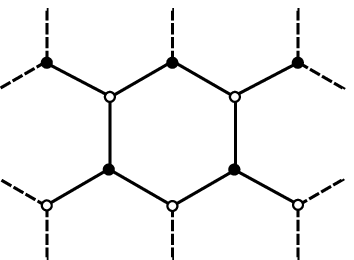}.
\end{eqnarray}

This is a fundamental difference between these tensor network states on bipartite and non-bipartite lattices.  On bipartite lattices, we have efficient representations of finite Abelian lattice gauge theories. On non-bipartite lattices we still have highly non-trivial spin-$S$ quantum states, but with reduced topological order.  
Another influence of the lattice geometry on the state, its coordination number, is revealed in the structure of reduced density matrices across different bipartite lattices.  When we consider all the sites around a vertex, we see correlations in the reduced density matrix consistent with the branching rules.  If we trace over just one degree of freedom, a product density matrix arises directly.
\begin{eqnarray}
\minigrpx{2.5cm}{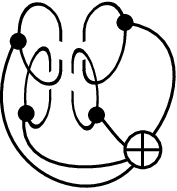}:\quad \minigrpx{2.5cm}{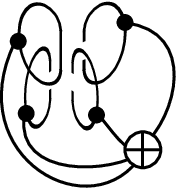} &= \minigrpx{2.5cm}{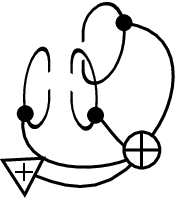}\nonumber\\
&= \minigrpx{2.5cm}{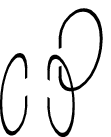}
\end{eqnarray}
Consequently, on a bipartite lattice correlations can only be revealed in the reduced density matrix if all $n$ spins around a vertex are considered.  In the case of non-bipartite lattices, tracing out one degree of freedom does not yield a a product density matrix, as the more complicated form of \eref{eq:S_top_kagome} suggests.

\section{Conclusion}\label{sec:conclusion}
We have adapted well known laws from modern algebra and extended techniques to a known tensor network representation of the $Z_2$ state, an archetypal topologically-ordered phase.   In particular we have demonstrated the existence of simple graphical rules by which the network may be contracted, allowing for the straightforward computation of physical quantities of relevance.  Via a natural generalisation in the on-site dimension of this network, we have found a class of tensor network states which display remarkably straightforward contraction properties as a consequence of the non-trivial algebraic properties of the tensor network components.
This is to be contrasted against tensor network states in general, for which efficient contraction is at best approximate. 
Furthermore, we have shown that this class forms the most efficient TNS representation of finite Abelian lattice gauge theories.

Our construction allows us to study the influence on the quantum state of varying the lattice geometry for a given tensor network.  The identical mathematical structures on all bipartite lattices signify identical physics for each case.  
Furthermore, these tensor networks also yield non-trivial quantum states on non-bipartite lattices, but only the $Z_2$ subgroups of the base group $G$ make a contribution to the topological entanglement entropy, in contrast to the bipartite case.  The nature of the algebraic contraction is such that a finite number of impurity regions can be handled, as the rest of the network can be contracted around them, leaving a small network to consider directly.  We have made use of this to calculate density matrices for string-like regions containing local perturbations, and shown how these can disentangle the region from the rest of the state.  We have also used this to calculate how a perturbations induce correlations into the ideal $Z_2$ state, which must be accounted for when calculating the topological entanglement entropy from the tensor network.

There are numerous directions for expanding this work, with some of the more immediate generalisations including the study of states based on non-Abelian and continuous groups, and classical lattice models based on networks of this variety.  More complex topological phases have tensor network representations, such as the double semion model and string-net states in general~\cite{PhysRevB.79.085118}.  These tensor networks involve double- and triple-line structures, but may also have internal structure enabling a similar set of local algebraic rewrite relations.

\ack
% All authors and co-authors are required to disclose any potential conflict of interest when submitting their article (e.g.~employment, consulting fees, research contracts, stock ownership, patent licenses, honoraria, advisory affiliations, etc). This information should be included in an acknowledgments section at the end of the manuscript (before the references section). All sources of financial support for the project must also be disclosed in the acknowledgments section. The name of the funding agency and the grant number should be given.
DJ and SRC thank the National Research Foundation and the Ministry of Education of Singapore for support.

\section*{References}

\bibliographystyle{sjd_general}
\bibliography{biblio}

\end{document}